%% file: main.tex
\documentclass[runningheads]{llncs}
\usepackage{standalone}
\usepackage{graphicx}
\usepackage{amsmath}
\usepackage{tabularx}
\usepackage{booktabs}
\usepackage[table]{xcolor}
\usepackage{multirow} 
\usepackage{color, colortbl}
\usepackage{todonotes}

\usepackage{nicefrac}

\usepackage{cclicenses}

\begin{document}
\title{Towards Greener Applications: Enabling Sustainable Cloud Native Applications Design \\ \large \texttt{preprint - submitted to CAiSE 2022} \ccby}
\titlerunning{Towards Greener Applications: Enabling Sustainable Design}
%
\author{Monica Vitali\inst{1}\orcidID{0000-0002-5258-1893} }
\authorrunning{M. Vitali}
%
\institute{Dipartimento di Elettronica, Informazione e Bioingegneria,
Politecnico di Milano, \\
\email{monica.vitali@polimi.it}\\}
\maketitle              
\begin{abstract}

Data centers energy demand is increasing. While a great deal of effort has been made to reduce the amount of CO$_2$ generated by large cloud providers, too little has been done from the application perspective. We claim that application developers can impact the environmental footprint by enhancing the application design with additional features. Following the proposed Sustainable Application Design Process (SADP), the application design is enriched with information that can be leveraged by cloud providers to manage application execution in an energy-aware manner.
This exploratory work aims to emphasize the awareness on the sustainability of applications by proposing a methodology for its evaluation.
To this end, we first suggest possible actions to enrich the application design towards sustainability, and finally describe how this additional information can be leveraged in the application workflow. We discuss the feasibility of our methodology by referring to existing tools and technologies capable of supporting the design features proposed in a production environment.

\end{abstract}
\input{sections/01_introduction.tex}
\input{sections/02_related_work.tex}
\input{sections/03_problem_description.tex}

\input{sections/04_maturitymodel_new.tex}
\input{sections/05_proposed_approach.tex}
\input{sections/06_validation}
\input{sections/07_conclusion.tex}

%
%
\bibliographystyle{splncs04}
\bibliography{sigproc}

\end{document}

%% file: sections/01_introduction.tex
\section{Introduction}
\label{sec:intro}
The last 10 years have seen an exponential growth in data centers energy demand. Recent studies have demonstrated that data centers are responsible for the 3\% of the global electricity supply and the 2\% of total greenhouse gas emissions\footnote{ \scriptsize  http://www.independent.co.uk/environment/global-warming-data-centres-to-consume-three-times-as-much-energy-in-next-decade-experts-warn-a6830086.html}. Efforts have been made, and are still ongoing, by leading cloud providers and IT companies, such as Facebook, Google, and Apple, towards environmental sustainability\footnote{\scriptsize  https://cloudscene.com/news/2016/12/going-green/}. To improve the efficiency and effectiveness of data centers, two complementary approaches can be adopted: (i) design efficient facilities~\cite{pierson2019datazero} and (ii) improve server utilization~\cite{pedram2012energy}. The first can be pursued by building data center facilities in cold locations, taking advantage of local renewable resources to power up the data center. The second strategy adopts server consolidation exploiting virtualization techniques. To date, most of the efforts have been addressed to the optimization of the Power Usage Effectiveness (PUE), a metric comparing the amount of energy consumed in a data center by the IT facilities and the overall energy consumed to power the whole data center (including cooling and uninterruptible power supply units). PUE gives hints about the energy efficiency of a data center but fails in expressing how efficiently IT resources are employed for running the applications~\cite{yuventi2013critical}.
Even if big companies have adopted both strategies, the overall carbon footprint of data centers has not changed significantly. While data centers are more efficient, the demand for data center services is increasing\footnote{\scriptsize  http://fortune.com/2016/06/27/data-center-energy-report/}
\cite{masanet2020recalibrating}. Computational demanding applications, including AI, machine learning, and Big Data analytics, consume a significant amount of energy, increased 300’000 times over the past 10 years~\cite{wired2020}\cite{lucivero2020big}. Therefore, the improvement in the efficiency of the infrastructure is followed by an increase in demand, making current efforts towards energy efficiency less relevant. 
At the same time, the architectural style of the applications is shifting from monolith to cloud native applications, designed to take advantage of the characteristics of cloud computing~\cite{kratzke2017understanding}~\cite{gannon2017cloud}. Cloud native applications are implemented through microservices: a large set of simple, single-function, and loosely coupled components interacting to provide the overall service. From the cloud provider perspective, these components are black boxes whose internal logic is neither declared nor modifiable, and whose interaction is mainly hidden. 

The main goal of the methodology proposed in this exploratory paper is to engage application designers in the path towards IT and IS sustainability. This paper focus the attention on the active role of applications in the energy footprint of IT and aims at increasing the awareness of application providers on the environmental footprint of their applications.  
The main contributions are:
\begin{itemize}
    \item a Sustainable Application Design Process (SADP), defining steps the designers need to perform to increase the sustainability of applications;
    \item a set of best-practices, guiding the application designers in improving the sustainability level of applications;
    \item insights on how SADP can be exploited in the application workflow.
\end{itemize}

The paper is organised as follows. Sect.~\ref{sec:sota} describes existing work. Sect.~\ref{sec:motivation} motivates the proposed approach with a running example. Sect.~\ref{sec:maturity} introduces SADP in more details. Sect.~\ref{sec:run-time} describes the employment of SADP at run-time. Sect.~\ref{sec:validation} validates the methodology mapping its steps to existing tools and technologies, while Sect.~\ref{sec:conclusion} summarizes the approach and outlines future developments.

%% file: sections/02_related_work.tex
\section{State of the Art}
\label{sec:sota}

The energy consumption in data centers has been taken into consideration for several years. Research has focused on the energy efficiency of data centers, with approaches related to the employment of renewable energy~\cite{goiri2013parasol} and the improvement of cooling efficiency~\cite{liu2012renewable}. The focus on cooling has been driven by adopting the Power Usage Efficiency (PUE) metric to measure data centers' energy efficiency, computed as the ratio of the amount of energy consumed for IT operations and the overall amount of energy. PUE has several limitations as it does not account for the type of energy used (brown or renewable) and the efficiency of IT operations~\cite{garrett2017green}. Some approaches have focused on IT resource management in data centers. In~\cite{surveyEnergyIJCIS}, approaches to green IT have been classified into three main categories: assessment, measurement, and improvement. Most of the approaches in improving energy efficiency are related to the infrastructure, exploiting the intermittent availability of renewable energy~\cite{thi2020game}\cite{9319812}. These involve operations such as server consolidation~\cite{pedram2012energy}\cite{gholipour2020novel}. When enacting consolidation, shut down policies must be taken into account to ensure a trade-off between energy efficiency and performance~\cite{benoit2018reducing}. To express non-functional requirements of applications, a Goal-Oriented Requirement Engineering approach can be used~\cite{horkoff2019goal}. Other approaches have considered several improvement actions, including consolidation, migration, CPU frequency and voltage scaling, and virtual machine resources vertical scaling~\cite{wajid2015achieving}\cite{vitali2015learning}\cite{stavrinides2019energy}. Some best practices have been suggested over the years to promote data center efficiency improvement, as the EU Code of Conduct~\cite{EUCodeConduct} and the Data Center Maturity Model~\cite{DCMM2011}. From an IT perspective, plenty of work have been investigating how to make data centers greener, while some limited attempts have been made to include sustainability in the application design of Green Information Systems~\cite{nowak2012pattern}\cite{cappiello2011business}\cite{vom2013green}\cite{loos2011green}. These approaches propose general principles and lacks from practical solutions. Some proposal aimed to estimate energy efficiency of specific applications in embedded systems~\cite{beziers2020annotating}, but cannot be applied in complex cloud infrastructures. From a design perspective, in \cite{schneider2019principles} more specific guidelines are provided focusing only on data mining. Current research mainly focuses on dynamic resource allocation and scheduling according to energy efficiency optimization~\cite{9369047}\cite{9338525}. All these techniques are not exploiting the differences between microservices and their interaction. Some efforts have been made to estimate the environmental footprint of applications. The CodeCarbon initiative~\cite{lottick2019energy} provides a tool for estimating CO$_2$ emissions the geographical location and the energy mix of the country in which the application is deployed. Power is also one of the metrics considered in~\cite{brondolin2020black}, providing black-box monitoring for multi-component applications. This is a first step for enabling energy-awareness in microservice-oriented applications but the workflow enhancement perspective is still missing. 

Cloud native applications empower organizations to design and execute scalable, loosely coupled, resilient, manageable, and observable applications~\cite{gannon2017cloud}. Cloud native adoption is increasing, thus we need to refer to this kind of model for future developments in cloud computing application management. These applications generate complex workflows due to the adoption of the microservice architectural style. A standard way to model the interaction between microservices is missing, even though this information is crucial to enhance their management. In \cite{valderas2020microservice}, microservice choreography is represented through BPMN fragments, thus exploiting a well-known process modeling notation for representing microservice interactions. However, the dynamicity of this interaction cannot be mapped in such an approach. To introduce dynamicity in microservice coordination, it is possible to enrich the model using the DMN specification \cite{omgDMN}, expressing business rules to define under which condition a task should be executed. An integration between BPMN and DMN is proposed in~\cite{hasic2018augmenting}\cite{neumann2019extending}, implementing the separation of concern between the process and decision model. 

This work focuses on enriching the design of cloud native applications and of their workflow in order to enable the sustainability of applications.

%% file: sections/03_problem_description.tex
\section{Motivating Scenario}
\label{sec:motivation}

In this section, we describe the main motivations driving this work and how the proposed approach is mapped in the current cloud scenario.

Our main goal is to raise attention towards sustainability of applications, and not only of the infrastructure in which they are deployed. In order to do so, we need to know the current state of the art in terms of cloud applications. For this reason, in this work we refer to cloud native applications, which are the current best practise for cloud applications. In fact, more and more developers are shifting towards the cloud native paradigm which requires to implement applications as a set of composite microservices, uncoupled and independent, interacting with each others through synchronous and asynchronous messages. A typical application is composed of dozens to hundreds of independent microservices, all implicitly cooperating to reach the overall goal of the organization implementing the application. However, \textbf{not all the components have the same relevance}, being some of them necessary for reaching the goal, while other just enriching the application with additional accessory functionalities that might increase the overall Quality of Experience (QoE) of the customer or the income of the service provider. Moreover, \textbf{ each microservice has different requirements} in terms of computational resources and different constraints in terms of Quality of Service (QoS). From this perspective, microservice based architectures are really effective since they allow the scale-out of the single components that are experiencing performance issues. Microservices can be sensitive to the context in which they are executed. Each microservice provides a specific functionality, however, the way in which this functionality is carried out might depend on the context of execution, and slightly different \textbf{alternative implementations of the same functionality can be provided}, i.e., with fail-over mechanisms. 

\begin{figure*}[!t]
\centering
\includegraphics[width=.55\textwidth]{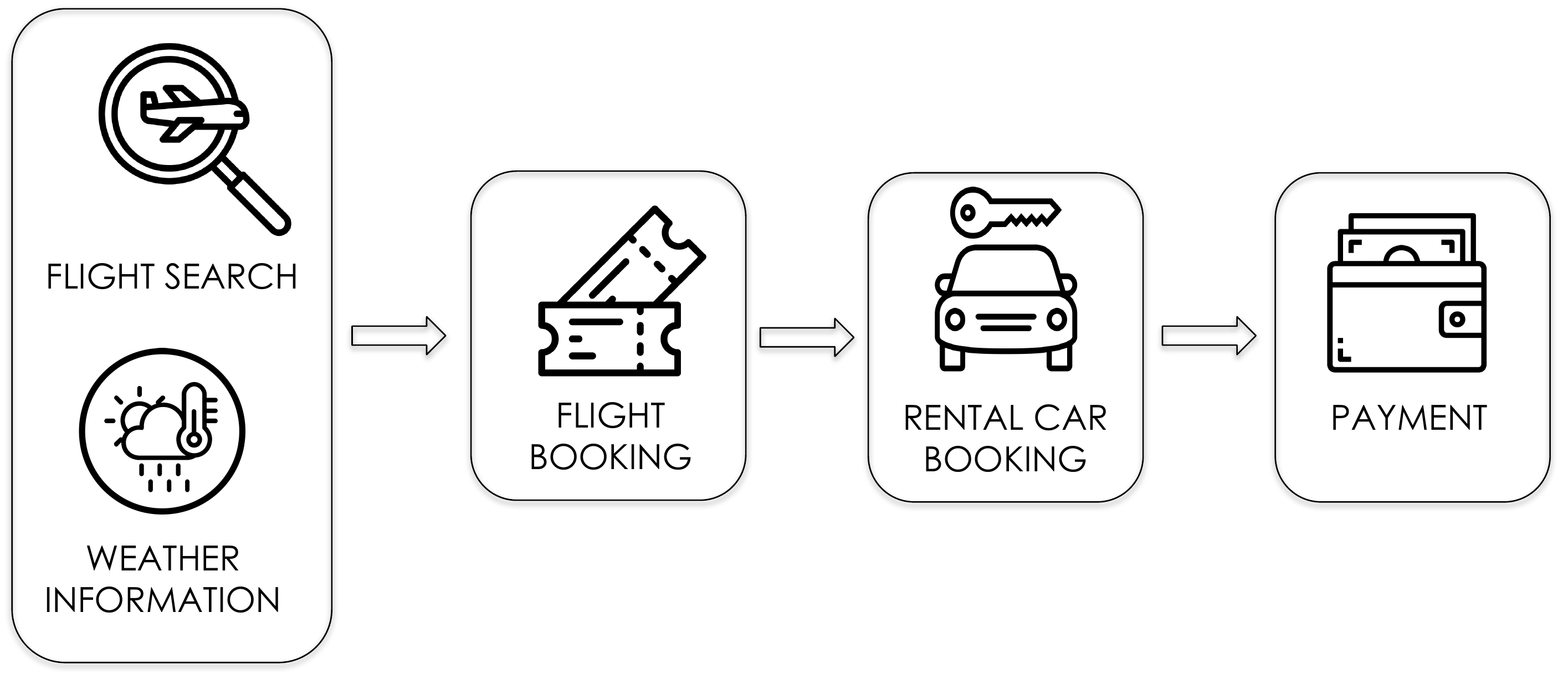} 
\caption{Microservices of a Flight Booking application}\label{fig:app}
\end{figure*}

To make the discussion more practical, let's introduce an example of application presenting the features we have just introduced. Let's consider a Flight Booking service allowing customers to check for specific itineraries and compare prices of several airlines. This service consists of several microservices (Fig.~\ref{fig:app}):

\begin{itemize}
    \item \textit{Flight Search}: it collects the itinerary request of the customers and returns a list of solutions obtained by querying the information systems of all the airlines. The solutions are ranked according to specific policies (e.g., price, duration, number of stops). This service can be provided with some variations according to the specific context of execution:
        \begin{itemize}
            \item \textbf{variation 1:} the information collected profiling the customer can be used to suggest routes or rank the results. This variation requires a recommendation engine to run in the background increasing the computational cost of the service while providing a better QoE to the customers;
            \item \textbf{variation 2:} results of recent searches by the same or other users can be reused. The actual query is executed only if the solution is selected. This might generate out of date results (e.g., not updated the cost of tickets for an airline) while reducing the computation time and cost.
        \end{itemize}
    \item \textit{Weather Information}: the search service can be complemented with information useful to the customers in selecting their itinerary. A weather service shows forecasts and statistics of temperatures and precipitations for the selected destination and dates that can be valuable for the customer.
    \item \textit{Flight Booking}: it is executed when a customer selects a solution after the search microservice. It includes all the activities related to the booking, including configurations (e.g., seat selection, baggage options) and the interaction with the airline's information system.
    \item \textit{Rental Car Booking}: additional services are proposed to the customer as the rental car booking. This service is provided by a partner but generates an income for the organization in case the customer books a vehicle. 
    \item \textit{Payment}: the payment service manages all the activities related to the payment of the selected flight solution.
\end{itemize}

Even in this very simple example, it is possible to see how some components of the application are mandatory (i.e., necessary to reach the goal of the application) while others are optional (i.e., contributing to the QoE but not affecting the overall goal). Examples of optional microservices are the \textit{Weather Information} and the \textit{Rental Car Booking} microservices. Moreover, the same functionality can be provided with some variations that might affect the QoS, the QoE, or the resource demand of the application, as demonstrated for the \textit{Flight Search} microservice. Finally, each microservice has different resources and QoS requirements. For instance, the \textit{Flight Search} might require some time to be executed (especially when querying all the airlines' systems) but results should be provided to the customers in a limited time to avoid that they turn to a competitor. On the contrary, a delay in showing the weather forecasts provided by the \textit{Weather Information} microservice is not problematic, although it is not desirable.

Most of these features, enclosing the complexity and dynamicity of cloud native applications, are hidden in configurations set by the application developer. These configurations are usually decided a priori and  rarely adapt with the context of execution. The cloud provider, who is the one in charge of managing most of the deployment aspects of the application, has no access to this information. Giving the provider the faculty of managing these configurations might severely improve the QoS of the application as well as its sustainability. Thus, it is important that the two stakeholders share relevant information for the effective management of the application.
The goal of this paper is to define which are the key features in the application design that can be exploited to improve the sustainability of the application. Secondly, it suggests how these features can be made explicit in the design of an application, and finally it discusses how to exploit them for improving the sustainability of the application workflow.

%% file: sections/04_maturitymodel_new.tex
\section{Sustainable Application Design Process}
\label{sec:maturity}

\begin{figure*}[!t]
\centering
\includegraphics[width=.7\textwidth]{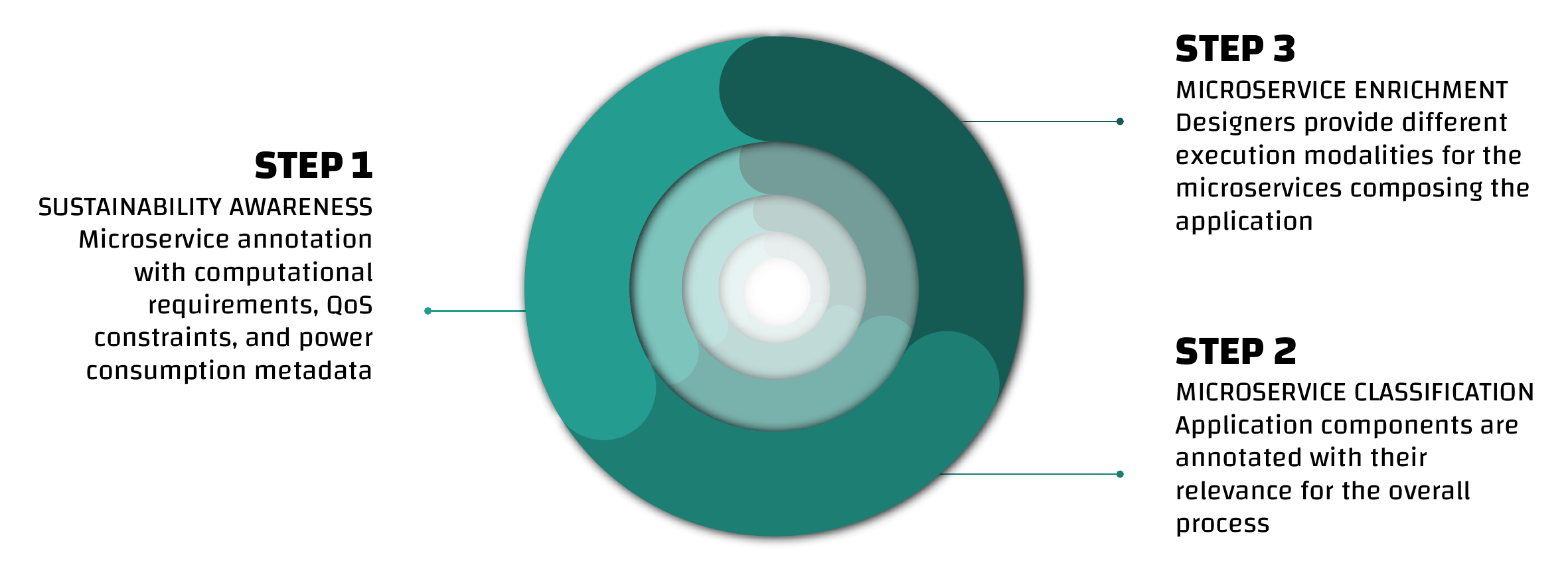} 
\caption{Steps of the SADP}\label{fig:samm}
\end{figure*}

The proposed Sustainable Application Design Process (SADP) aims at supporting designers in the achievement of sustainable microservice based applications. SADP focuses on an application perspective by providing goals and directions for designing sustainable applications.
The steps in Fig.~\ref{fig:samm} have been identified.

\noindent \textbf{Step 0} The application designer is not putting effort in designing sustainable applications. Level 0 is the current state of the art. An application can be described as a graph $A = \{M, E\}$
where $M$ is the set of microservices composing the application and $E$ is the set of edges connecting the microservices. 

\noindent \textbf{Step 1} is a first step towards sustainable applications. It consists in adding basic sustainability-aware information to the application design. The goal is to enrich the application model with additional information that can drive the deployment decisions of the cloud provider when the single components are deployed. This information is not only related to the energy aspect, but also to functional and QoS requirements. The set $R$ contains the specification of all the requirements that can be expressed for an application as $R = \{F, Q, S\}$ 
where: $F$ is the set of functional requirements regarding the amount of required resources (e.g., the size of the VM required for deploying that component); $Q$ is the set of quality related non-functional requirements (e.g., response time or throughput constraints), and $S$ is the set of sustainability related non-functional requirements (e.g., the estimated power consumption for executing the microservice). Estimating power consumption of a microservice is not trivial, since providing such information is not straightforward. However, from a simple profiling activity it is possible to estimate the computational power required by a specific VM or container configuration. Models available in the state of the art, transforming computational power in energy consumption, can then be exploited \cite{brondolin_deep-mon_2018}.
For each microservice $m_i \in M$, a set of requirements $R_i \subseteq R$ can be expressed. A score can be assigned to the application measuring the extent to which the application design implements SADP's Step 1 best practices. Given the set $R$ of expressible requirements, the score is assigned according to the coverage of the annotation:
\begin{equation}
    A.score_1 = \nicefrac{\sum_i |R_i|}{|R| |M|} \; \; \; A.score_1 \in [0,1]
\end{equation}
\noindent where $|R_i|$ is the number of requirements expressed for a microservice $m_i \in M$, and $|R| |M|$ is the total amount of requirements that can be expressed for $A$.

\noindent \textbf{Step 2} The application designer provides information on which components of the application are mandatory and which are optional: $M_M \cup M_O \subseteq M$ 
where $M_M$ is the set of mandatory microservices and $M_O$ is the set of optional microservices.
Considering the flight reservation application example in Fig.~\ref{fig:app}, the rental car service can provide additional value for the customer and additional revenue for the application owner, but it is not a key component of the main application. If information about the relevance of each component is provided at design time, this information can be exploited to decide when to execute or when to skip a component according to the execution context. This information is relevant not only for the sustainability of the application (e.g., skip an optional component when renewable power source is not available), but also for QoS (e.g., avoid optional components when the application is experiencing response time or latency issues). Two possible approaches can be used to implement Step 2: i) all components require annotations stating if they are necessary or not for the overall workflow, ii) only optional components are annotated, while mandatory components are not explicitly identified. In the first case, a score can be assigned according to the amount of components explicitly annotated in the workflow:
\begin{equation}
    A.score_2 = \nicefrac{(|M_O| + |M_M|)}{|M|}   \; \; \; A.score_2 \in [0,1]   
\end{equation}
\noindent However, this approach is more time consuming. In the second approach, instead, a refined score cannot be associated, since it is not possible to differentiate between not annotated and implicitly annotated components. Thus, for the second approach, the score can only be 0 if no annotation is provided, and 1 if at least one component has been annotated as optional:
\begin{equation}
    A.score_2 = \small \begin{cases} 0 \; \; \; if \; M_O \cup M_M = \emptyset \\
    1 \; \; \; otherwise
    \end{cases}
\end{equation}

\noindent \textbf{Step 3} It consists in the enrichment of the microservices composing the application through the definition of different modalities of execution for each component. Alternative execution modalities can be defined for a single microservice and the one to enact might depend on the current context. Here, three different execution modalities (or versions) are proposed: $V = \{ N, HP, LP\} $.
\begin{itemize}
    \item Normal (N): this is the basic execution modality of the microservice;
    \item High-Performance (HP): the performance of the microservice is stressed so that the QoS and/or QoE of the customer is maximized at the cost of a higher computational resource demand and power comsumption;
    \item Low-Power (LP): it is a simplified version of the microservice aiming at reducing the amount of energy consumed. To guarantee the performance level, some sub-activities are skipped or executed differently. As an example, in the flight reservation application, the ``Search flight'' component might take into account the previous activities of the customer in ranking the results to provide a better QoE. However, this has an additional computational cost that can be reduced by ignoring specific user information. 
\end{itemize}
\noindent For each microservice, both optional and mandatory, it is possible to define several versions: $ m_i.V = \{ m_i.v\} \: \: | \; m_i \in M \;, \; v \in V \;, \; |m_i.V| \leq |V|$; 
where $m_i$ is a microservice of the application, $m_i.V$ is the set or modalities defined for $m_i$, and $m_i.v$ is a specific modality of the microservice (Normal, Low-Power, or High-Performance).
A score can be assigned according to the number of alternative versions provided compared to the overall amount of required versions:
\begin{equation}
    A.score_3 = \nicefrac{\sum_i{|m_i.V|}}{|M| |V|}  \; \; \; A.score_3 \in [0,1]
\end{equation}
Different modalities correspond to different power consumption, QoS, and QoE, thus a trade-off is to be considered when selecting the proper modality.
The overall SADP approach and the suggested best practices are summarized in Fig.~\ref{fig:samm}. The four steps are incremental refinements. At each step, additional information is provided by the application designer and, at the same time, the complexity in the management of the application increases together with the degree of freedom of how to execute the application. The process is incremental: at each loop, the designer can focus on a single component or on a subset of components, enriching the model step by step.


%% file: sections/05_proposed_approach.tex
\section{Sustainable Workflow Design with SADP}
\label{sec:run-time}

SADP provides a methodology for improving and evaluating the sustainability in the design of a cloud-native application, enriching each component with additional details and metadata. This section describes how it is possible to exploit these features designing sustainable workflows. 

Designing a sustainable workflow requires to select the best configuration for all the components of the application according to the context, aiming at improving energy efficiency, cost, and QoS.
The composition of the application will dynamically change selecting a different pool of microservices according to the context of execution. Exploiting all the features added in the process design described in Sect.~\ref{sec:maturity}, the workflow can be executed in different modalities:
\begin{itemize}
    \item \textbf{Normal execution} is the typical behaviour of an application; all the microservices composing the application are executed when a request arrives. 
    \item \textbf{Basic execution} only a subset of the tasks composing the application are executed. Not mandatory tasks are skipped to reduce the energy consumption or to improve the QoS. It is supported by the Step 2 of the SADP. 
    \item \textbf{Low-power execution} tasks composing the application are executed using a low power modality if available (e.g., using editorial content instead of computing personalized content);
    \item \textbf{High-performance execution} tasks composing the application are executed using a performance enhanced modality if available (e.g., providing additional personalizations and functionalities to improve the QoE).
\end{itemize}

In order to enable the \textbf{Low-power execution} and the \textbf{High-performance execution}, all the microservices, or part of them, have to be designed according to the Step 3 of the SADP: different interfaces to execute the microservice are provided for each supported modality. These workflow execution modalities are not exclusive, in fact they can be combined together. As an example, if a Low-power execution is activated together with a Basic execution, all the optional tasks are skipped, while regular tasks are executed using their low-power version if available.
Two different approaches can be applied for the enactment of the workflow modalities just described:
\begin{itemize}
    \item \textbf{All in}: the execution modality is globally selected for all the components of the application. With the Low-power execution, all the microservices providing this behaviour are executed in this modality without any distinction. Similarly, with the Basic execution, all the optional tasks are skipped.
    \item \textbf{Optimized selection}: the decision about which modality to activate is performed at the microservice level. The execution is optimized by selecting the best combination for reaching the overall desired performance.
\end{itemize}

While the second approach enables a fine grained optimisation, it increases the complexity of the workflow design and management, which makes necessary the introduction off a complex decision process at run-time. Thus, the workflow design has to be enriched with decision points that are able to provide the logic for deciding the execution modality according to the actual context of execution. 

%% file: sections/06_validation.tex
\section{Validation}
\label{sec:validation}

This section demonstrates the feasibility of the proposed methodology given the current technological stack and the cloud native landscape. 


\input{sections/03a_application_model.tex}

\subsection{Designing Sustainable Workflows}\label{subsec:workflow}

\begin{figure*}[!t]
\centering
\includegraphics[width=.99\textwidth]{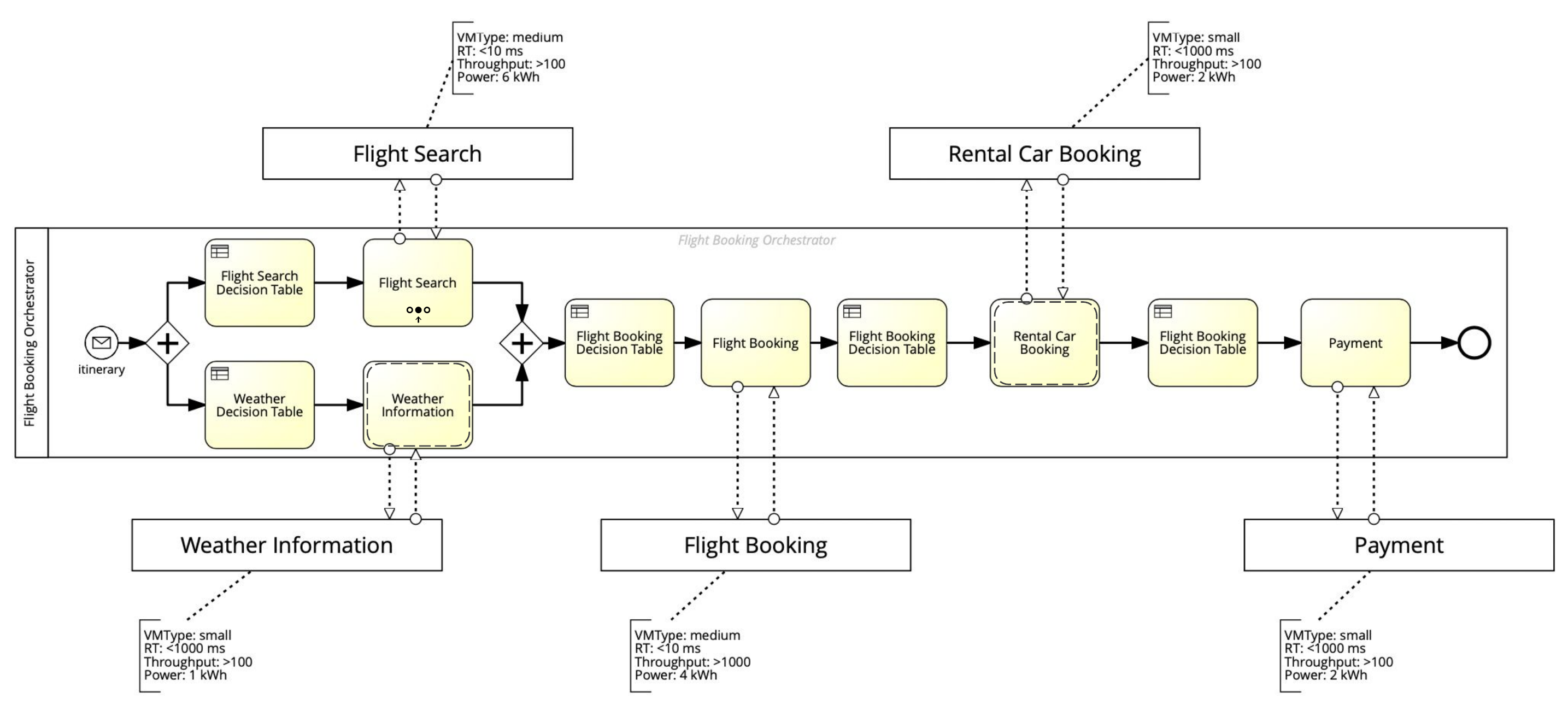} 
\caption{Flight Booking Workflow Orchestrator Model}\label{fig:orchestrator}
\end{figure*}

In order to exploit the features provided by SADP and enable the different execution modalities introduced in Sec.~\ref{sec:run-time}, the process model is enriched with business rules. Business rules enable to express conditions on the execution of an activity in a business process. They hide the complexity of the logic preventing the model to become too complex. In this case, business rules are included to define the execution modality of each task according to the values of a set of contextual variables collected by the monitoring system of the application and each of its microservices. Each business rule is linked to a DMN model~\cite{omgDMN}, which makes explicit the involved variables and the decision tables containing the business rules used in the process. A decision table contains rules in the form if-then, where the condition is based on the value of the contextual variables. An example for the Flight Booking application is shown in Fig.~\ref{fig:orchestrator}. Each component is preceded by a business rule, linked to a DMN model, defining the modality of execution for the following microservice. Examples of business rules are:
\begin{itemize}\small
    \item \textit{Rental Car Booking}: \texttt{if response\_time > 1000 ms then skip};
    \item \textit{Flight Search}: \texttt{if power > 5 kW then low-power execution}.
\end{itemize}

These are very simple rules, but a general rule can combine different variables and more complex reasoning. However, this paper demonstrated that the logic behind the execution modality can be included in the application design.

Business rules will depend on the workflow modality enactment approach. In the case of the \textbf{All in} approach, an execution modality is selected at the process level, thus a single DMN model is referred by the business rules for establishing if the normal, basic, low-power, or high-performance execution should be enacted. For instance, in case of poor energy efficiency, the low power execution and the basic execution modalities can be selected, resulting in the execution of the \textit{Flight Search} microservice in low power mode and in the skipping of both the \textit{Weather Information} and the \textit{Rental Car Booking} microservices. On the contrary, in the case of \textbf{Optimized execution} a custom modality will be selected for each microservice, thus requiring more complex DMN models.

\subsection{Feasibility and open challenges}\label{subsec:run-time}

In the previous sections, the proposed approach has been validated by demonstrating that modeling tools are available to capture all the features introduced in the proposed methodology. We will now look at how these features can be supported in the cloud native ecosystem.
The outcome of the design step will be an annotated and enriched BPMN model containing all the relevant information for the management of sustainability-aware applications. BPMN models can be translated in machine-readable formats like XML or JSON, including annotations and task types that will enrich the description of each task. It is thus possible to translate the model into a set of features that can be exploited by an orchestrator. As an example, a BPMN-driven microservice coordination approach is provided by some existing tools in the state of the art. For instance, Camunda\footnote{ \scriptsize Camunda: Microservices and BPMN: https://bit.ly/3pxYPbS} and Zeebee\footnote{\scriptsize  Zebee: A Workflow Engine for Microservices Orchestration: https://zeebe.io} are workflow engines supporting BPMN and DMN for designing workflow orchestrations. The other aspect is related to the management of dynamic execution modalities for microservices. The concept of alternative execution paths for microservices, activated according to the current context is supported by the state of the art technological stack. When a microservice of the pipeline experiences performance issues it can be replaced by a different one until the problem is not solved (fail-over management). Even if these approaches are traditionally related to the QoS management \cite{aldwyan2019latency}, their mechanism can be easily adapted to other goals such as sustainability.

SADP is based on the assumption that monitoring the energy consumption (and even better the CO$_2$ emissions) of each single component of an application is possible. At the moment this information is not provided by any cloud provider natively. However, several initiatives are proposing solutions for estimating both energy and emissions at the application level\footnote{\scriptsize Cloud Jewels: https://codeascraft.com/2020/04/23/cloud-jewels-estimating-kwh-in-the-cloud/ \\ Cloud Carbon Footprint: https://www.cloudcarbonfootprint.org/ \\ CodeCarbon: https://mlco2.github.io/codecarbon/visualize.html}. This shows a growing interest in the topic that is expected to result in additional improvement of existing third party tools and in the involvement of cloud providers in the process. Another critical aspect is the increased complexity in the application design, requiring to define several versions for each microservice. From a design perspective, the proposed approach is incremental, leaving the application owner the right to decide to what extent to refine the existing application model. The initial effort can be limited on the most affecting components and extended only in future steps to other components. From a management perspective, costs will probably increase. Thus, the problem is how to motivate application providers to invest in sustainable applications. Existing studies have demonstrated how sustainability can become a strategic value for both organizations and their customers thanks to a proper awareness of the impact on the environment~\cite{schneider2019principles}. 





%% file: sections/03a_application_model.tex
\subsection{Designing Sustainable Applications}
\label{subsec:design-time}

\begin{figure*}[!t]
\centering
\includegraphics[width=.9\textwidth]{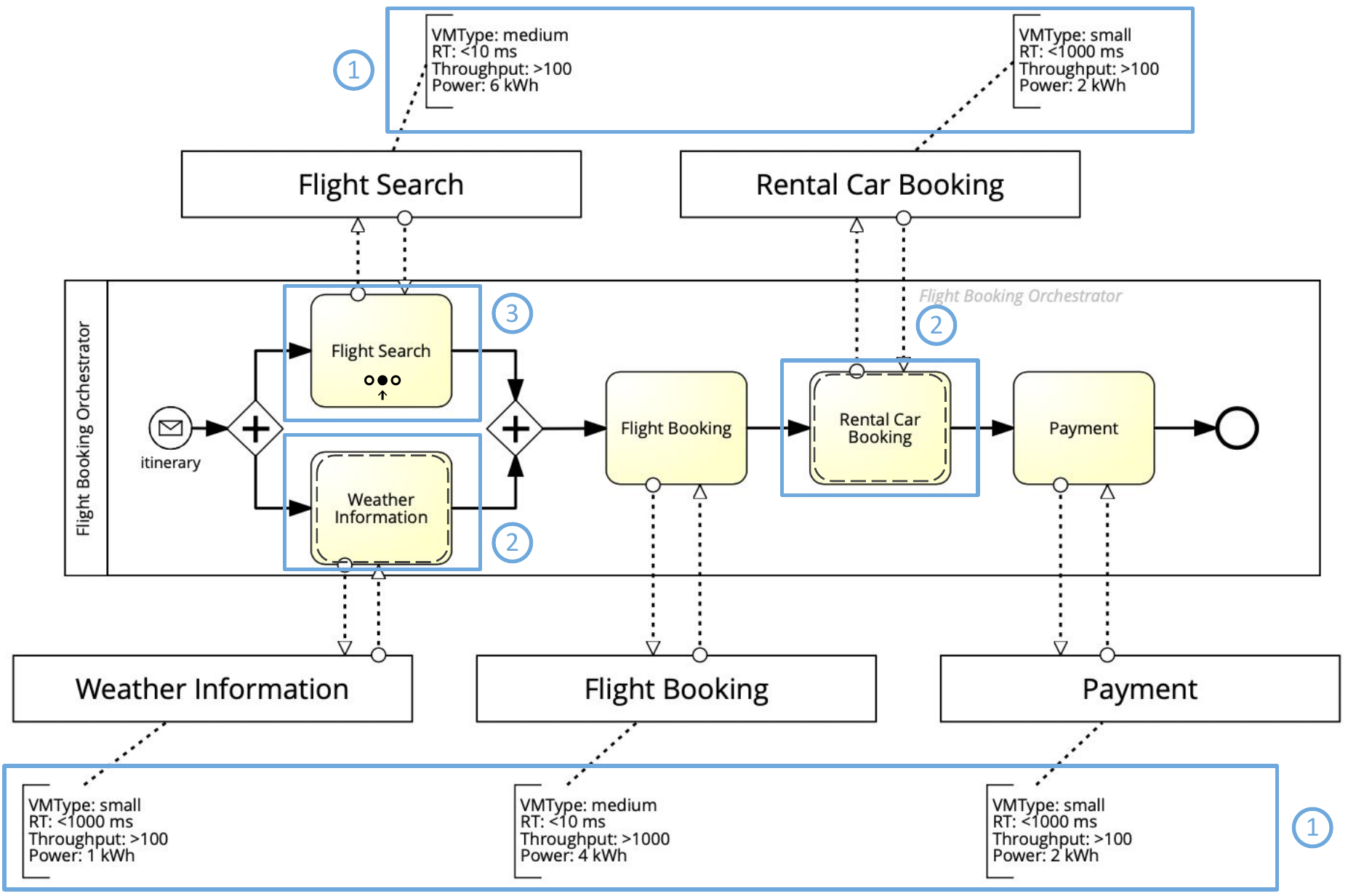}
\caption{BPMN representation of the Flight Booking application redesigned with SADP}\label{fig:bpmn_flightAll}\label{fig:bpmn_flight}
\end{figure*}

In Sect.~\ref{sec:maturity} we introduced SADP, identifying the features to obtain sustainability-aware applications. This section proposes some solutions for designing sustainable applications based on microservices. Here we propose to represent the application and its components using the BPMN notation \cite{omg2011bpmn}, a standard used to represent both internal processes of an organization as well as collaborations and orchestrations between different organizations. In this work, applications are a workflow of operations performed by different microservices, that might belong to different organizations. Following the representation adopted in \cite{valderas2020microservice}, the application can be represented as a set of microservices orchestrated by a general process defining their order of execution. In this way, we can model the interaction between different microservices through message events generated during the execution of a microservice or at its end, triggering the execution of another microservice or influencing its activities. As an example, the BPMN model of the Flight Booking application is represented in Fig.~\ref{fig:bpmn_flight}. The details of the logic of the single microservices are hidden since they are not relevant for the discussion. The interaction is made explicit through the orchestrator process. 

Following the steps presented in Sect.~\ref{sec:maturity}, the process model can be enriched with metadata regarding the functional and non functional requirements of each microservice. In order to do so, it is possible to use annotations. Annotations are already defined in BPMN and they are used to add additional information on the task execution. Here, annotations are used to add details on the microservice functional and non functional requirements (e.g., required computational resources, QoS constraints), execution cost, power consumption, and eventual reward associated to the microservice execution. These meta-data are useful to predict which will be the outcome of executing a specific microservice in the process workflow. The outcome of Step 1 applied to the flight booking application is shown in Fig.~\ref{fig:bpmn_flightAll}. Metadata are modeled with the annotation artifact of BPMN using a semi-structured notation. In order to obtain a consistent annotation, each aspect is represented as an attribute-value pair and the set of allowed attributes and values is predefined. The score is assigned according to the coverage of the current annotation. As an example, assuming that only 4 attributes can be expressed, the process in Fig.~\ref{fig:bpmn_flightAll} has a Step 1 score of 100\% since all the attributes are defined for all the components.
The Step 2 process representation is enriched with the information about the optional tasks. This can be implemented enriching the BPMN notation with an additional type of task represented with an internal dashed border, indicating that a component is optional for the execution. For instance, in the flight booking application, the \textit{Weather Information} and the \textit{Rental Car Booking} microservices are indicated as optional, as shown in Fig.~\ref{fig:bpmn_flightAll}. This is because they don't affect the reach of the final goal of the process (booking a flight) even if their execution improves the QoE providing the customer with additional information and functionalities. In case these functionalities are provided by third parties, skipping their execution might reduce the income for the organization. As an example, the flight booking service might have an agreement with the rental car company consisting in a percentage gain of the overall rental contracts stipulated through the booking service. Skipping this tasks results in the lost of the possible revenue originated by a car rental. This step is not associated with a refined score, since it is not possible for anyone to know which are the necessary steps of the process but for the owner of the process. 
At Step 3, the process is enriched with alternative execution modalities. It is represented enriching the BPMN notation with a type identifier for the task representing the call to the component in the orchestrator. As an example, in Fig.~\ref{fig:bpmn_flightAll} the \textit{Flight Search} microservice is identified as a component with alternative execution modes. In the example, five microservices are involved but only one has been refined with the alternative versions (both low power and performance enhanced). Thus, the score will be 20\%. 


%% file: sections/07_conclusion.tex
\section{Conclusion}
\label{sec:conclusion}

This paper introduced the SADP methodology to support the design of sustainable applications. SADP proposes an incremental process, defining levels of sustainability for application design and several enrichment that can be exploited to improve the application energy-efficiency while maintaining the QoS. The paper demonstrates how the proposed solution can be supported by existing tools and technologies. The SADP methodology is a first step towards energy-aware application management and aims at engaging application owners in the path towards sustainable IT. However, there are still several challenges to motivate application developers' effort to make their applications more sustainable, but we believe these benefits can compensate the additional costs and efforts in the long term.
In future work we will provide a framework supporting SADP and implement a prototype exploiting all the introduced features for sustainable applications design and execution. We will also consider an additional step: how to optimize the deployment of microservices in a heterogeneous fog environment to minimize the energy footprint of applications at run-time.